\documentclass[preprint]{revtex4-1} 
\usepackage[latin1]{inputenc}
\usepackage{graphicx}
\usepackage{bm}
\usepackage{amsmath}
\usepackage{xcolor}
\usepackage{array}
\usepackage{braket}
\usepackage{microtype}
\usepackage{appendix}

\renewcommand{\baselinestretch}{1.0}
\begin{document}

\renewcommand{\baselinestretch}{1.0}

\title{Quantifying interference in multipartite quantum systems}

\author{Rejane Alves de Brito$^{1}$ and Bert\'ulio de Lima Bernardo$^{1,2}$}\email{bertulio.fisica@gmail.com}

\affiliation{$^{1}$Departamento de F\'isica, Universidade Federal da Para\'iba, 58051-900 Jo\~ao Pessoa, PB, Brazil\\
$^{2}$Departamento de F\'isica, Universidade Federal de Campina Grande, Caixa Postal 10071, 58109-970 Campina Grande-PB, Brazil}

\begin{abstract}

The characterization of quantum correlations is crucial to the development of new quantum technologies and to understand how dramatically quantum theory departs from classical physics. Here we systematically study single- and multiparticle interference patterns produced by general two- and three-qubit systems. From this we establish on phenomenological grounds a new type of quantum correlation for these systems, which we name {\it quantum interference}, deriving some quantifiers that are given explicitly in terms of the density matrix elements of the complete system. By using these quantifiers, we show that, contrary to our expectations, entanglement is not a required property for a composite quantum system to manifest multiparticle interference.

\end{abstract}

\maketitle


\section{Introduction}

The concept of wave-particle duality, commonly described as the ability of a quantum particle to produce interference, is a central ingredient of quantum theory, which is absent in our classical intuition of the physical world. In fact, as famously stated by Feynman \cite{feynman}, this is ``the mystery'' manifested by microscopic particles, ``which is impossible, absolutely impossible, to explain in any classical way, and which has in it the heart of quantum mechanics''. Perhaps, the best illustration of wave-particle duality, now understood as a consequence of the quantum superposition principle, is given by the double-slit experiment. This experiment was originally presented by Young in the early 1800s to ascertain the wave properties of light and has become widely used to understand many fundamental aspects of quantum mechanics since the inception of the theory, e.g. the complementarity principle \cite{bohr}. In this experiment, a beam of particles impinges on a mask with two closely spaced slits through which some of them can pass to have their position detected by a sensitive screen placed on the opposite side. For the case in which there is no information about which slit each of the particles traverses, the detection screen exhibits an interference pattern, therefore, revealing the wave behavior \cite{bert1}. The first experimental realization of the double-slit experiment on the molecular level was presented in the 1960s using electrons \cite{jonsson}, later conducted with $C_{60}$ \cite{arndt}, as well as with larger molecules \cite{eib}. However, only a few years ago this experiment could be realized in full agreement with Feynman's idea \cite{bach}.                  

In the late 1980s, a remarkable new class of experiments expanding the successful category of single-particle interferometers was inaugurated; the so-called multiparticle interferometers. Ref. \cite{green} provides a comprehensive review on this topic. At that time, the new interferometers brought about an alternative way of studying quantum phenomena which are richer and even more intriguing than those resulting from quantum superposition, evidenced in single-particle systems. The experimental realizations were first obtained using entangled photons \cite{rarity,ghosh,ribeiro}, and more than a decade later were performed using the internal states of four ions in a single trap \cite{sac}. Above all, it was claimed that multiparticle interferometers can provide measurable outcomes that are associated with the very nature of quantum entanglement. In this case, the signature of entanglement was recognized to be a surprising interference effect that takes place if one monitors the arrival positions of the particles composing the system in coincidence \cite{hsz1,hsz2}. In contrast, no single-particle interference is detectable with this configuration. This behavior, although surprising, is to some extent intuitive if we recall that quantum superposition gives rise to wave-particle duality, and that entanglement was the name given by Schr\"{o}dinger to quantum superposition in a many-particle system \cite{schro}. As a matter of fact, one of the main lessons that we have learned from the longstanding studies with interferometers was that the fundamental concepts of quantum superposition and entanglement are the responsible for the emergence of single- and multiparticle interference, respectively.                      

Here, we study the single- and multiparticle interference behavior of bipartite and tripartite systems when each of the particles is submitted to a double-slit interferometer. From the analysis of the bipartite case, we derive a formula to quantify the amount of two-particle interference, which is shown to be a quantum correlation different from Bell nonlocality \cite{brunner}, entanglement\cite{vedral} and discord \cite{modi}. Similarly, we show how to quantify three-particle interference for the important case of general three-qubit systems. In this last case, we identify two different classes of three-particle interference, one represented by $GHZ$-like states and the other by $W$-like states. Nevertheless, contrary to what has been believed so far \cite{jwp,jwp2}, we find that it is possible for a composite quantum system to produce multiparticle interference without being entangled. The article is organized as follows: In Sec. II we give a clear definition of two-particle interference relying on a scheme in which both particles are individually submitted to a double-slit interferometer. Based on the interference patterns manifested in this thought experiment, we derive an expression that quantifies quantum interference for general two-qubit system and provide some key examples to compare with the results of other nonclassical correlations. In Sec. III, we extend the quantum interference study to the important case of three-qubit systems. Again, the system is analyzed under the perspective that each particle experiences a double-slit apparatus. The validity of the derived three-particle interference quantifier is also illustrated with some examples of pure and mixed states. Conclusions and remarks are given in
Sec. VI.

\section{Two-qubit case}

In this section we want to quantify interference for a general two-qubit system in terms of the interference properties that the subsystems produce upon individual and joint observation. To do so, we consider a thought experiment which consists in a source which emits pairs of particles, $a$ and $b$, in opposite directions so that each particle is submitted to a double-slit experiment to be further detected on a screen which marks the detection positions. The scheme, which was previously studied by one of the authors in the context of decoherence theory \cite{bert}, is shown in Fig. 1. Here, we call $\ket{A_{1}}$ and $\ket{A_{2}}$ the states of particle $a$ when it passes through the upper and lower slits on the right, respectively. Similarly, we denote by $\ket{B_{1}}$ and $\ket{B_{2}}$ the respective states of particle $b$. If we use the four states $\ket{A_{i}}\ket{B_{j}}$, with $i,j=1,2$, as a basis to describe this bipartite system, the general state is described by the density operator 
\begin{eqnarray} \label{rho}
\begin{aligned}
\hat{\rho}=
&\rho_{11}\ket{A_1}\ket{B_1}\bra{A_1}\bra{B_1}+\rho_{12}\ket{A_1}\ket{B_1}\bra{A_1}\bra{B_
2}
+\rho_{13}\ket{A_1}\ket{B_1}\bra{A_2}\bra{B_1}+\\
&\rho_{14}\ket{A_1}\ket{B_1}\bra{A_2}\bra{B_2}+\rho_{21}\ket{A_1}\ket{B_2}\bra{A_1}\bra{B
_1}+
\rho_{22}\ket{A_1}\ket{B_2}\bra{A_1}\bra{B_2}+\\
&\rho_{23}\ket{A_1}\ket{B_2}\bra{A_2}\bra{B_1}+\rho_{24}\ket{A_1}\ket{B_2}\bra{A_2}\bra{B
_2}+
\rho_{31}\ket{A_2}\ket{B_1}\bra{A_1}\bra{B_1}+\\
&\rho_{32}\ket{A_2}\ket{B_1}\bra{A_1}\bra{B_2}+\rho_{33}\ket{A_2}\ket{B_1}\bra{A_2}\bra{B
_1}+
\rho_{43}\ket{A_2}\ket{B_1}\bra{A_2}\bra{B_2}+\\
&\rho_{41}\ket{A_2}\ket{B_2}\bra{A_1}\bra{B_1}+\rho_{42}\ket{A_2}\ket{B_2}\bra{A_1}\bra{B
_2}+
\rho_{43}\ket{A_2}\ket{B_2}\bra{A_2}\bra{B_1}+\\
&\rho_{44}\ket{A_2}\ket{B_2}\bra{A_2}\bra{B_2}
,
\end{aligned}
\end{eqnarray}
whose density matrix has 16 entries: $\rho_{ij}$ with $i,j=1,2,3$ and $4$.

Now, if we are interested in obtaining information about the joint probability of detecting particle $a$ at a point $z_{A}$ on the left screen  and particle $b$ at $z_{B}$ on the right screen in coincidence, this is given by the relation   
\begin{eqnarray} \label{sigma}
\rho(z_A,z_B)\equiv \bra{z_A}\bra{z_B} \hat{\rho} \ket{z_B}\ket{z_A},
\end{eqnarray}
where $\ket{z_{A}}$ and $\ket{z_{B}}$ are the respective position eigenstates.
Let us consider that after passing through a slit, the wavefunction associated with the emerging particle is
spherical. This assumption allows us to write the wavefunctions
\begin{eqnarray}
\psi_{Al}(r_{Al})=\braket{r_{Al}|A_l}=\dfrac{e^{ikr_{Al}}}{r_{Al}}
\end{eqnarray}
and
\begin{eqnarray}
\psi_{Bl}(r_{Bl})=\braket{r_{Bl}|B_l}=\dfrac{e^{ikr_{Bl}}}{r_{Bl}},
\end{eqnarray}
for the particles on the left and on the right side, respectively. Here, $k$ represents the wavenumber, and $r_{(A, B)l}$ the distances
from the slits to the detection points, with $l = 1,2$, as seen in Fig. 1.

\begin{figure}[ht]
\centerline{\includegraphics[width=11cm]{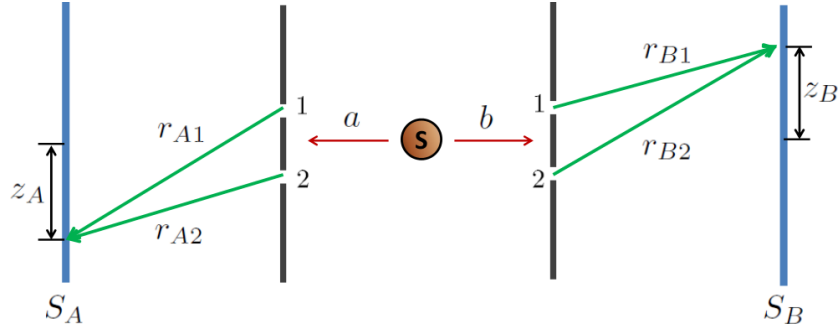}}
\caption{(Color online) Schematic illustration for the two-qubit system discussed in the text. A source $S$ generates a pair of particles $a$ and $b$, which are submitted to the left and right double-slit apparatuses, respectively. After passing through the double-slit stage, the particles are detected on the screens $S_{A}$ and $S_{B}$, which permanently mark their position as $z_{A}$ and $z_{B}$. The observable manifestation of single- and two-particle interference in this apparatus is used to quantify interference for general two-qubit systems.}
\label{setup}
\end{figure}

In the regime in which the distance between the slits is much smaller than the distance between the double-slit apparatus and the screen, the Fraunhofer diffraction limit is valid \cite{bert,gott}, such that we have
$r_{A(1,2)}\approx L \mp\theta z_A$ and $r_{B(1,2)}\approx L \mp\theta z_B$, with $L$ and $\theta$ defined in Fig. 2. Taking all this information into consideration, together with Eq. \eqref {rho}, it turns out that the joint probability density will have the following form:
\begin{eqnarray} \label{rho2}
\begin{aligned}
\rho(z_A,z_B)=& \dfrac{\rho_{11}}{L^3[L-2\theta(z_A+z_B)]}+\dfrac{\rho_{22}}{L^3[L-
2\theta(z_A-z_B)]}+
\dfrac{\rho_{33}}{L^3[L+2\theta(z_A-z_B)]}+ \\
&\dfrac{\rho_{44}}{L^3[L+2\theta(z_A+z_B)]}+\dfrac{2}{L^4}\bigg[Re({\rho_{23}e^{-
2ik\theta(z_A-z_B)}})+
Re({\rho_{14}}e^{-2ik\theta(z_A+z_B)})+ \\
& Re({\rho_{24}}e^{-2ik\theta z_A})+Re({\rho_{13}}e^{2ik\theta z_A})
+Re({\rho_{12}}e^{2ik\theta z_B})+
Re({\rho_{34}}e^{-2ik\theta z_B}) \bigg].
\end{aligned}
\end{eqnarray}

After some algebra, we find that the joint probability density becomes
\begin{eqnarray} \label{rho22}
\begin{aligned}
\rho(z_A,z_B)=& \dfrac{\rho_{11}}{L^3[L-2\theta(z_A+z_B)]}+\dfrac{\rho_{22}}{L^3[L-
2\theta(z_A-z_B)]}+
\dfrac{\rho_{33}}{L^3[L+2\theta(z_A-z_B)]}+ \\
&\dfrac{\rho_{44}}{L^3[L+2\theta(z_A+z_B)]}+\dfrac{2}{L^4}\bigg[ [R_{23}+R_{14}]\cos(2k\theta z_A)\cos(2k\theta z_B)+\\
&[R_{23}-R_{14}]\sin(2k\theta z_A)\sin(2k\theta z_B) +[I_{23}+I_{14}]\sin(2k\theta z_A)\cos(2k\theta z_B)+\\
&[I_{14}-I_{23}]\cos(2k\theta z_A) \sin(2k\theta z_B)+[R_{13}+R_{24}]\cos(2k\theta z_A)+\\
&[R_{12}+R_{34}]\cos(2k\theta z_B)+[I_{24}-I_{13}]\sin(2k\theta z_A)+[I_{34}-I_{12}]\sin(2k\theta z_B)\bigg],
\end{aligned}
\end{eqnarray}
where we used the notation $\rho_{ij}=R_{ij}+iI_{ij}$ for the entries of the density matrix, with $i$ as the imaginary unit, i.e., $R_{ij}$ and $I_{ij}$ are the real and imaginary parts of $\rho_{ij}$, respectively. Observe that the first four terms in Eq.~(\ref{rho22}) (outside brackets) represent the probability density of detecting particles $a$ and $b$, respectively, at $z_{A}$ and $z_{B}$ for the cases in which there is a complete information about the path they have taken, i.e., the slit they have traversed. These terms do not contribute with any type of interference effects. The first four terms inside brackets quantifies the existence of oscillation in the coincidence detection rate (CDR) of particles $a$ and $b$. Finally, the last four terms inside brackets are the responsible for the single-particle interference effects. That is, the existence of these terms gives rise to individual spatial interference patterns on the detection screens.

\begin{figure}[ht]
\centerline{\includegraphics[width=10cm]{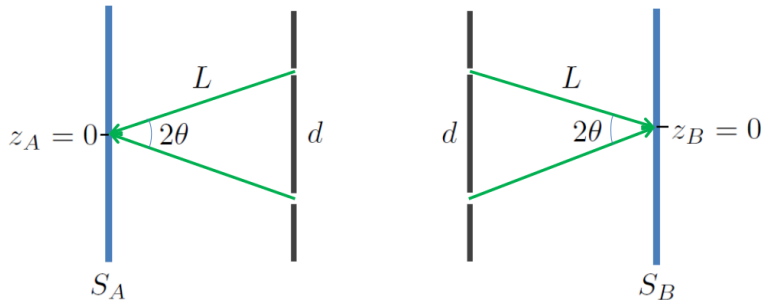}}
\caption{(Color online) Schematic diagram of the apparatus presented in Fig. 1. The angle $\theta$, as well as the distances $L$ and $d$, are useful parameters to describe the probability density of particles detected on the screens $S_{A}$ and $S_{B}$.}
\end{figure} 

A genuine oscillation in the CDR of particles $a$ and $b$ at the distant screens is a phenomenon that we expect only if these particles have a nonclassical correlation. Therefore, at this point we argue that the ``detectability'' of this type of oscillation is directly linked to the amount of {\it quantum interference} that these two particles have. Later, we show that this correlation is {\it not} entanglement, as widely accepted in the literature of multiparticle interference. Having thus described the problem, we shall seek a proper quantifier of multiparticle interference, starting from this simplest case of two qubits. However, it is important to envisage that, even if particles $a$ and $b$ present only single-particle interference on their respective screens, these two independent interference patterns exert an influence on the measured oscillation in the CDR obtained when the data of the two screens are considered together. In this form, we must keep in mind that genuine two-particle interference only occurs if, after subtracting the single-particle oscillation contributions from the two-particle ones, the oscillation in the CDR still remains. In other words, this residual oscillatory effect is the signature of genuine two-particle interference, which implies the existence of quantum-mechanical correlations. 

In order to be more precise, observe, for example, from Eq.~(\ref{rho22}) that if the sum $2(R_{23}+R_{14})$ and the product $4[(R_{13}+R_{24})(R_{12}+R_{34})]$ are nonzero, both will contribute with an interference mode of the type $\cos(2k \theta z_{A})\cos(2k \theta z_{B})$. However, the first term embodies the contributions to this mode of {\it both} genuine two-particle interference and the combined single-particle interferences fringes, whereas the second term represents a contribution {\it only} of the combined single-particle interference fringes formed on each screen. Therefore, if these two terms are equal, it is because there is no genuine two-particle interference with this mode. On the other hand, if the first term contribution is larger than the second, the system manifests genuine two-particle interference, i.e., quantum correlations. Similarly, three other independent analyses can be made relating the contribution of single- and two-particle interference. If the terms $2(R_{23}-R_{14})$ and  $4[(I_{24}-I_{13})(I_{34}+I_{12})]$ are nonzero, they contribute with an interference of the type $\sin(2k \theta z_{A})\sin(2k \theta z_{B})$; if the terms $2(I_{23}+I_{14})$ and  $4[(I_{24}-I_{13})(R_{12}+R_{34})]$ are nonzero, both contribute with a $\sin(2k \theta z_{A})\cos(2k \theta z_{B})$ interference mode; and if the terms $2(I_{14}-I_{23})$ and  $4[(R_{13}+R_{24})(I_{34}+I_{12})]$ survive, they will contribute with an interference of the type $\cos(2k \theta z_{A})\sin(2k \theta z_{B})$. It is important to observe that these four types of two-particle interference compose a basis for a two-dimensional Fourier series, which can generate any sinusoidal function $f(z_{A},z_{B})$, with periodicity $\ell=\pi/k \theta$ both in $z_{A}$ and $z_{B}$, 

\begin{eqnarray} \label{f1}
\begin{aligned}
f(z_A,z_B)=& A \cos\left(\frac{2 \pi }{\ell}z_{A}\right)\cos\left(\frac{2 \pi }{\ell}z_{B}\right)+ B\sin\left(\frac{2 \pi }{\ell}z_{A}\right)\sin\left(\frac{2 \pi }{\ell}z_{B}\right)+\\
& C \sin\left(\frac{2 \pi }{\ell}z_{A}\right)\cos\left(\frac{2 \pi }{\ell}z_{B}\right)+ D\cos\left(\frac{2 \pi }{\ell}z_{A}\right)\sin\left(\frac{2 \pi }{\ell}z_{B}\right),
\end{aligned}
\end{eqnarray}
with 
\begin{equation}
    A=4/\ell^{2}\int^{\ell/2}_{-\ell/2} \int^{\ell/2}_{-\ell/2} f(z_A,z_B) \cos\left(\frac{2 \pi }{\ell}z_{A}\right)\cos\left(\frac{2 \pi }{\ell}z_{B}\right) d z_{A} d z_{B},
\end{equation}
\begin{equation}
    B=4/\ell^{2}\int^{\ell/2}_{-\ell/2} \int^{\ell/2}_{-\ell/2} f(z_A,z_B) \sin\left(\frac{2 \pi }{\ell}z_{A}\right)\sin\left(\frac{2 \pi }{\ell}z_{B}\right) d z_{A} d z_{B},
\end{equation}
\begin{equation}
    C=4/\ell^{2}\int^{\ell/2}_{-\ell/2} \int^{\ell/2}_{-\ell/2} f(z_A,z_B) \sin\left(\frac{2 \pi }{\ell}z_{A}\right)\cos\left(\frac{2 \pi }{\ell}z_{B}\right) d z_{A} d z_{B},
\end{equation}
\begin{equation}
    D=4/\ell^{2}\int^{\ell/2}_{-\ell/2} \int^{\ell/2}_{-\ell/2} f(z_A,z_B) \cos\left(\frac{2 \pi }{\ell}z_{A}\right)\sin\left(\frac{2 \pi }{\ell}z_{B}\right) d z_{A} d z_{B}.
\end{equation}
As a result, it signifies that the four types of interference are linearly independent (LI), such that their influence can be analyzed separately. 

Based on the arguments above, in what follows we shall establish our quantum interference quantifier for a pair of qubits built upon the oscillations in the CDR that they are capable to produce. In this respect, we mathematically define our quantifier with basis on the imbalance between the two- and single-particle interference contributions to the CDR for each of the four LI oscillatory modes. If this imbalance is null, the distant particles do not manifest any detectable oscillation of this type, and consequently the state has no (multiparticle) quantum interference. This feature indicates the complete absence of quantum correlations between the two particles. On the other hand, if there exists some nonzero oscillation in the CDR of the particles, it is because the state of the particles has some amount of two-particle interference. Under this phenomenological definition, our two-qubit quantum interference quantifier assumes the following form:      
\begin{eqnarray}  \label{quant}
\begin{aligned}
 \mathcal{I}_{2}^{(2)}(\hat{\rho}) \equiv  2\bigg\{&|(R_{23}+R_{14})^2-4[(R_{13}+R_{24})(R_{12}+R_{34})]^2|\\
  + &|(R_{23}-R_{14})^2-4[(I_{24}-I_{13})(I_{34}-I_{12})]^2|\\
  + &|(I_{23}+I_{14})^2-4[(I_{24}-I_{13})(R_{12}+R_{34})]^2|\\
  + &|(I_{14}-I_{23})^2-4[I_{34}-I_{12})(R_{13}+R_{24})]^2|\bigg\},
\end{aligned}
\end{eqnarray}
which is given explicitly in terms of the entries of the density matrix, $\rho_{ij}=R_{ij}+iI_{ij}$. The subscript and superscript 2 in $ \mathcal{I}_{2}^{(2)}(\hat{\rho})$ stand for the number of parties and the dimension of the Hilbert space of each party, respectively. The four {\it absolute value} terms in Eq.~(\ref{quant}) represent the imbalance between the {\it square} of the two- and single-particle contributions to each of the four LI modes: $\cos(2k \theta z_{A})\cos(2k \theta z_{B})$, $\sin(2k \theta z_{A})\sin(2k \theta z_{B})$, $\sin(2k \theta z_{A})\cos(2k \theta z_{B})$ and $\cos(2k \theta z_{A})\sin(2k \theta z_{B})$, respectively \cite{comment}. We also included a multiplicative (normalization) factor $1/2$ in order for the maximally entangled Bell states provide $\mathcal{I}_{2}^{(2)}(\hat{\rho}) = 1$.       

Let us now present some key examples of the application of the interference quantifier of Eq.~(\ref{quant}). For the sake of clarity, in the following examples we will use the simpler computational basis states: $\ket{A_{1}}\ket{B_{1}} \rightarrow \ket{00}$, $\ket{A_{1}}\ket{B_{2}} \rightarrow \ket{01}$, $\ket{A_{2}}\ket{B_{1}} \rightarrow \ket{10}$ and $\ket{A_{2}}\ket{B_{2}} \rightarrow \ket{11}$. Accordingly, the pure two-qubit states $\ket {\Phi(\theta,\phi)}=\cos \theta\ket{00}+e^{i\phi}\sin \theta \ket{11}$ and $\ket {\Psi(\theta,\phi)}=\cos \theta\ket{01}+e^{i\phi}\sin \theta \ket{10}$ provide the result $\mathcal{I}_{2}^{(2)}=\sin^2(2\theta)$. This result is satisfactory since it is independent of $\phi$, and $\mathcal{I}_{2}^{(2)}=1$ for the four maximally entangled Bell states, which are obtained when $\theta = \pi/4$ and $\theta = 3\pi/4$ (for $\phi=0$), and $\mathcal{I}_{2}^{(2)}=0$ for the separable states that emerge when $\theta = 0$ and $\theta = \pi/2$. Another key example is the case of an arbitrary separable pure state $\ket{\psi} = \ket{\psi_{A}}\ket{\psi_{B}}$, with $\ket {\psi_{A}}=\cos (\theta_1 /2)\ket{0}+e^{i\phi_1}\sin (\theta_1 /2)\ket{1}$ and $\ket{\psi_{B}}=\cos (\theta_2 /2)\ket{0}+e^{i\phi_2}\sin (\theta_2 /2) \ket{1}$. For this case, after some calculations, Eq.~(\ref{quant}) yields $\mathcal{I}_{2}^{(2)}=0$, which is the expected result for an arbitrary separable pure state.

Another interesting fact to observe is that Eq.~(\ref{quant}) does not depend on the diagonal terms of the density matrix. Thus, for any mixed state of the type 
\begin{equation}
\quad
\hat{\rho}=\begin{pmatrix} 
\rho_{11} & 0 &0 & 0\\
0 & \rho_{22} & 0& 0\\
0 & 0 & \rho_{33}& 0\\
0 & 0 & 0& \rho_{44}\\
\end{pmatrix},
\quad  
\end{equation}
we have $\mathcal{I}_{2}^{(2)}=0$, as expected. From the fact that the quantifier is necessarily zero if the density matrix has all off-diagonal elements null, one might be tempted to consider some connections between this concept and that of quantum coherence, whose quantifiers also have this property \cite{baumg,bu}. To properly address this question, we must first remember that quantum coherence is not a type of quantum correlation. In fact, this can be clearly understood if we recall that while the former property can be associated to single quantum systems, the latter is necessarily linked to composite ones. Second, it is well-known that coherence is a basis-dependent definition \cite{baumg,bu}, and, more similar to the approaches of entanglement and discord, our idea of quantum interference is basis-independent. This can be seen, for example, from the fact that all separable states provide $\mathcal{I}_{2}^{(2)}=0$, and all Bell states provide $\mathcal{I}_{2}^{(2)}=1$, as shown above, which means that local unitary operations do not change the amount quantum interference for these states. Investigations with other pure and mixed states also lead us to this conclusion. Yet, we believe it is worthwhile to seek for interconnections between the notions of quantum coherence and the multiparticle interference presented here, but these concepts are definitely {\it not} the same.

We now consider as an example a special case of two-qubit mixed states, the Werner state. This state, which plays a central role in quantum information theory, is defined as a mixture of a singlet state, $\ket{\psi^{(-)}} = 1/\sqrt{2}(\ket{01} - \ket{01})$, and completely
depolarized noise \cite{werner}:
\begin{equation}
\label{werner}
 \hat{\rho}_{W}= p \ket{\psi^{(-)}}\bra{\psi^{(-)}} + \frac{1-p}{4} \hat{{\bf 1}},
\end{equation}
with $0 \leq p \leq 1 $. Werner demonstrated that this state is entangled only if $p > 1/3$. Another remarkable result about this state is that quantum discord, another measure of nonclassical correlations, is nonzero whenever $p > 0$ \cite{ollivier,hend}. As a consequence, this result showed that there exists quantumness in some separable states. In Fig. 3 we show the behavior of quantum interference (given by the definition of Eq.~(\ref{quant})), entanglement and discord as a function of the parameter $p$.  

\begin{figure}[ht]
\centerline{\includegraphics[width=10.2cm]{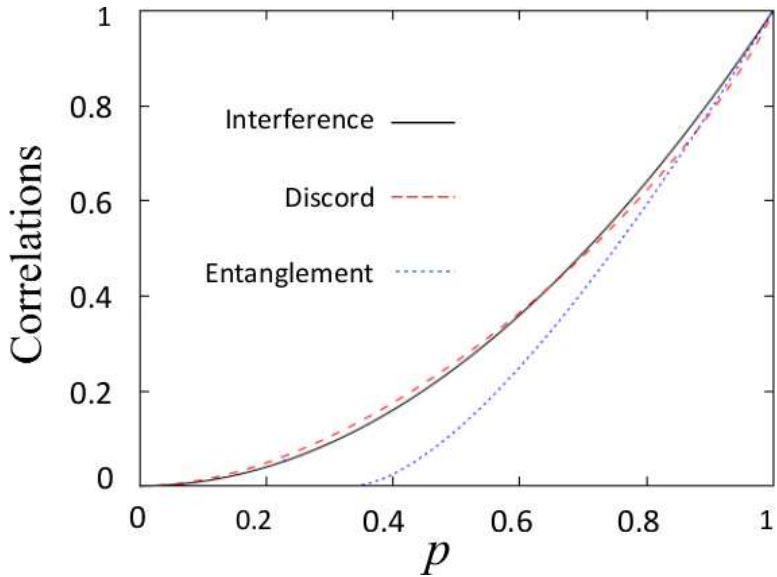}}
\caption{(Color online) Graphs of the quantum interference $\mathcal{I}_2^{(2)}(\hat{\rho}_{W})$, discord, and entanglement of formation for the two-qubit Werner state in Eq.~(\ref{werner}) as a function of $p$. As can be seen, both interference and discord are nonzero in the interval $0 \leq p \leq 1/3$, in which the state is separable. All three correlations reach unit when $p=1$.}
\end{figure} 
From the quantum interference curve, we can extract another remarkable result based on the study of Werner states: it is possible to have two-particle interference in the absence of entanglement. In fact, the separable Werner states given by $0 < p \leq 1/3$ are able to produce two-particle interference. This finding contradicts the widely accepted idea that multiparticle interference demands entanglement \cite{green,jwp,jwp2}. Furthermore, we also have that states with nonzero quantum interference do not necessarily violates a Bell inequality. Indeed, the state of Eq.~(\ref{werner}) only violates the CHSH inequality if $p > 1/\sqrt{2} \approx 0.707$ \cite{brunner}. In the next section we extend the study of quantification of interference to the three-qubit case.

\section{Three-qubit case}

In the same spirit of the previous section, we now want to quantify interference for a general three-qubit system by means of the interference properties that the subsystems manifest upon individual and joint observations. In the present case, we consider a {\it gedanken} experiment consisting in a source that creates simultaneously three particles $a$, $b$ and $c$, which are individually sent towards one of three double-slit apparatuses to be further detected on the screens $S_{A}$, $S_{B}$ and $S_{C}$, as shown in Fig. 4. Similar to the two-qubit case, the quantum states of the particles according to the slit they traverse are denoted by $\ket{A_{1}}$ and $\ket{A_{2}}$ for particle $a$; $\ket{B_{1}}$ and $\ket{B_{2}}$ for particle $b$; and $\ket{C_{1}}$ and $\ket{C_{2}}$ for particle $c$.         

\begin{figure}[ht]
\centerline{\includegraphics[width=10.2cm]{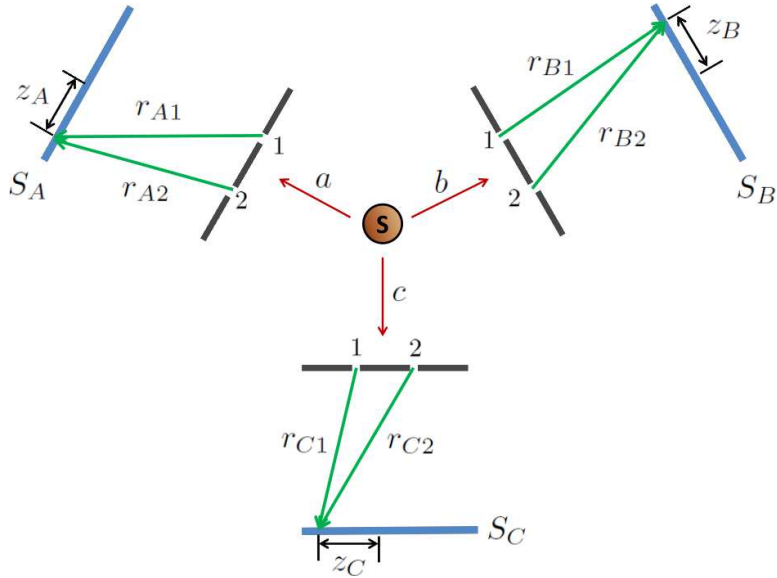}}
\caption{(Color online) Schematic illustration of the three-qubit system discussed in the text. A source $S$ creates three particles $a$, $b$ and $c$, which are individually submitted to a corresponding double-slit apparatus. After, the particles are detected on the screens $S_{A}$, $S_{B}$ and $S_{C}$, which mark their position at the points $z_{A}$, $z_{B}$ and $z_{C}$, respectively. By investigating the single-, two- and three-particle interference behavior that they produce, information about the tripartite quantum interference can be obtained.}
\end{figure} 

By choosing the eight product states $\ket{A_{i}}\ket{B_{j}}\ket{C_{k}}$, with $i,j,k=1,2$, as the basis states for representing the general tripartite states of the particles, we have that the density operator of the system is written as follows: 

\begin{eqnarray} \label{rho3}
\begin{aligned}
\rho &= \rho_{11}\ket{A_1}\ket{B_1}\ket{C_1}\bra{A_1}\bra{B_1}\bra{C_1} + \rho_{22}\ket{A_1}\ket{B_1}\ket{C_2}\bra{A_1}\bra{B_1}\bra{C_2} \\  &+\rho_{33}\ket{A_1}\ket{B_2}\ket{C_1}\bra{A_1}\bra{B_2}\bra{C_1} + \rho_{44}\ket{A_1}\ket{B_2}\ket{C_2}\bra{A_1}\bra{B_2}\bra{C_2} \\
&+ \rho_{55}\ket{A_2}\ket{B_1}\ket{C_1}\bra{A_2}\bra{B_1}\bra{C_1} + \rho_{66}\ket{A_2}\ket{B_1}\ket{C_2}\bra{A_2}\bra{B_1}\bra{C_2} \\
&+ \rho_{77}\ket{A_2}\ket{B_2}\ket{C_1}\bra{A_2}\bra{B_2}\bra{C_1} + \rho_{88}\ket{A_2}\ket{B_2}\ket{C_2}\bra{A_2}\bra{B_2}\bra{C_2} \\
&+\rho_{12}\ket{A_1}\ket{B_1}\ket{C_1}\bra{A_1}\bra{B_1}\bra{C_2}
+\rho_{21}\ket{A_1}\ket{B_1}\ket{C_2}\bra{A_1}\bra{B_1}\bra{C_1} \\
& +\rho_{13}\ket{A_1}\ket{B_1}\ket{C_1}\bra{A_1}\bra{B_2}\bra{C_1}
+\rho_{31}\ket{A_1}\ket{B_2}\ket{C_1}\bra{A_1}\bra{B_1}\bra{C_1}\\
& +\rho_{14}\ket{A_1}\ket{B_1}\ket{C_1}\bra{A_1}\bra{B_2}\bra{C_2}
+\rho_{41}\ket{A_1}\ket{B_2}\ket{C_2}\bra{A_1}\bra{B_1}\bra{C_1}\\
& +\rho_{15}\ket{A_1}\ket{B_1}\ket{C_1}\bra{A_2}\bra{B_1}\bra{C_1}
+\rho_{51}\ket{A_2}\ket{B_1}\ket{C_1}\bra{A_1}\bra{B_1}\bra{C_1}\\
& +\rho_{16}\ket{A_1}\ket{B_1}\ket{C_1}\bra{A_2}\bra{B_1}\bra{C_2}
+\rho_{61}\ket{A_2}\ket{B_1}\ket{C_2}\bra{A_1}\bra{B_1}\bra{C_1}\\
& +\rho_{17}\ket{A_1}\ket{B_1}\ket{C_1}\bra{A_2}\bra{B_2}\bra{C_1}
+\rho_{71}\ket{A_2}\ket{B_2}\ket{C_1}\bra{A_1}\bra{B_1}\bra{C_1}\\
& +\rho_{18}\ket{A_1}\ket{B_1}\ket{C_1}\bra{A_2}\bra{B_2}\bra{C_2}
+\rho_{81}\ket{A_2}\ket{B_2}\ket{C_2}\bra{A_1}\bra{B_1}\bra{C_1}\\
& +\rho_{23}\ket{A_1}\ket{B_1}\ket{C_2}\bra{A_1}\bra{B_2}\bra{C_1}
+\rho_{32}\ket{A_1}\ket{B_2}\ket{C_1}\bra{A_1}\bra{B_1}\bra{C_2}\\
& +\rho_{24}\ket{A_1}\ket{B_1}\ket{C_2}\bra{A_1}\bra{B_2}\bra{C_2}
+\rho_{42}\ket{A_1}\ket{B_2}\ket{C_2}\bra{A_1}\bra{B_1}\bra{C_2}\\
& +\rho_{25}\ket{A_1}\ket{B_1}\ket{C_2}\bra{A_1}\bra{B_1}\bra{C_1}
+\rho_{52}\ket{A_1}\ket{B_1}\ket{C_1}\bra{A_1}\bra{B_1}\bra{C_2}\\
& +\rho_{26}\ket{A_1}\ket{B_1}\ket{C_2}\bra{A_2}\bra{B_1}\bra{C_2}
+\rho_{62}\ket{A_2}\ket{B_1}\ket{C_2}\bra{A_1}\bra{B_1}\bra{C_2}\\
& +\rho_{27}\ket{A_1}\ket{B_1}\ket{C_2}\bra{A_2}\bra{B_2}\bra{C_1}
+\rho_{72}\ket{A_2}\ket{B_2}\ket{C_1}\bra{A_1}\bra{B_1}\bra{C_2}\\
& +\rho_{28}\ket{A_1}\ket{B_1}\ket{C_2}\bra{A_2}\bra{B_2}\bra{C_2}
+\rho_{82}\ket{A_2}\ket{B_2}\ket{C_2}\bra{A_1}\bra{B_1}\bra{C_2}\\
& +\rho_{34}\ket{A_1}\ket{B_2}\ket{C_1}\bra{A_1}\bra{B_2}\bra{C_2}
+\rho_{43}\ket{A_1}\ket{B_2}\ket{C_2}\bra{A_1}\bra{B_2}\bra{C_1}\\
& +\rho_{35}\ket{A_1}\ket{B_2}\ket{C_1}\bra{A_2}\bra{B_1}\bra{C_1}
+\rho_{53}\ket{A_2}\ket{B_1}\ket{C_1}\bra{A_1}\bra{B_2}\bra{C_1}\\
& +\rho_{36}\ket{A_1}\ket{B_2}\ket{C_1}\bra{A_2}\bra{B_1}\bra{C_2}
+\rho_{63}\ket{A_2}\ket{B_1}\ket{C_2}\bra{A_1}\bra{B_2}\bra{C_1}\\
& +\rho_{37}\ket{A_1}\ket{B_2}\ket{C_1}\bra{A_2}\bra{B_2}\bra{C_1}
+\rho_{73}\ket{A_2}\ket{B_2}\ket{C_1}\bra{A_1}\bra{B_2}\bra{C_1}\\
& +\rho_{38}\ket{A_1}\ket{B_2}\ket{C_1}\bra{A_2}\bra{B_2}\bra{C_2}
+\rho_{83}\ket{A_2}\ket{B_2}\ket{C_2}\bra{A_1}\bra{B_2}\bra{C_1}\\
& +\rho_{45}\ket{A_1}\ket{B_2}\ket{C_2}\bra{A_2}\bra{B_1}\bra{C_1}
+\rho_{54}\ket{A_2}\ket{B_1}\ket{C_1}\bra{A_1}\bra{B_2}\bra{C_2} \\
& +\rho_{46}\ket{A_1}\ket{B_2}\ket{C_2}\bra{A_2}\bra{B_1}\bra{C_2}
+\rho_{64}\ket{A_2}\ket{B_1}\ket{C_2}\bra{A_1}\bra{B_2}\bra{C_2}\\
& +\rho_{47}\ket{A_1}\ket{B_2}\ket{C_2}\bra{A_2}\bra{B_2}\bra{C_1}
+\rho_{74}\ket{A_2}\ket{B_2}\ket{C_1}\bra{A_1}\bra{B_2}\bra{C_2}\\
& +\rho_{48}\ket{A_1}\ket{B_2}\ket{C_2}\bra{A_2}\bra{B_1}\bra{C_2}
+\rho_{84}\ket{A_2}\ket{B_1}\ket{C_2}\bra{A_1}\bra{B_2}\bra{C_2}\\
& +\rho_{56}\ket{A_2}\ket{B_1}\ket{C_1}\bra{A_2}\bra{B_1}\bra{C_2}
+\rho_{65}\ket{A_2}\ket{B_1}\ket{C_2}\bra{A_2}\bra{B_1}\bra{C_1}\\
& +\rho_{57}\ket{A_2}\ket{B_1}\ket{C_1}\bra{A_2}\bra{B_2}\bra{C_1}
+\rho_{75}\ket{A_2}\ket{B_2}\ket{C_1}\bra{A_2}\bra{B_1}\bra{C_1}\\
& +\rho_{58}\ket{A_2}\ket{B_1}\ket{C_1}\bra{A_2}\bra{B_2}\bra{C_2}
+\rho_{85}\ket{A_2}\ket{B_2}\ket{C_2}\bra{A_2}\bra{B_1}\bra{C_1}\\
& +\rho_{67}\ket{A_2}\ket{B_1}\ket{C_2}\bra{A_2}\bra{B_2}\bra{C_1}
+\rho_{76}\ket{A_2}\ket{B_2}\ket{C_1}\bra{A_2}\bra{B_1}\bra{C_2}\\
& +\rho_{68}\ket{A_2}\ket{B_1}\ket{C_2}\bra{A_2}\bra{B_2}\bra{C_2}
+\rho_{86}\ket{A_2}\ket{B_2}\ket{C_2}\bra{A_2}\bra{B_1}\bra{C_2}\\
& +\rho_{78}\ket{A_2}\ket{B_2}\ket{C_1}\bra{A_2}\bra{B_2}\bra{C_2}
+\rho_{87}\ket{A_2}\ket{B_2}\ket{C_2}\bra{A_2}\bra{B_2}\bra{C_1},
\end{aligned}
\end{eqnarray}
which has 64 entries: $\rho_{ij}$ with $i$ and $j$ varying from 1 to 8.

In this case, we have that the joint probability of measuring particles $a$, $b$ and $c$ at the points $z_{A}$, $z_{B}$ and $z_{C}$ in coincidence is given by
\begin{eqnarray} \label{sigma3}
\rho(z_{A},z_{B},z_{C})\equiv \bra{z_A}\bra{z_B}\bra{z_{C}} \hat{\rho} \ket{z_{C}}\ket{z_B}\ket{z_A}.
\end{eqnarray}
By following the same steps of the previous section, if we assume that the wavefunctions of the particles after emerging from the slits are spherical, and the geometry of all three double-slit apparatuses are such that the Fraunhofer diffraction limit is applicable, we find that the joint probability density is given by:    

\begin{eqnarray} \label{rho31}
\begin{aligned}
\rho(z_A, z_B, z_C) = &\dfrac{\rho_{11}}{L^5[L-2\theta(z_A+z_B+z_C)]}+
\dfrac{\rho_{22}}{L^5[L+2\theta(-z_A+z_B+z_C)]} \\
&+\dfrac{\rho_{33}}{{L^5[L+2\theta(-z_A+z_B-z_C)]}}
+\dfrac{\rho_{44}}{{L^5[L+2\theta(-z_A+z_B+z_C)]}}\\
&+\dfrac{\rho_{55}}{{L^5[L+2\theta(z_A-z_B-z_C)]}}
+\dfrac{\rho_{66}}{{L^5[L+2\theta(z_A-z_B+z_C)]}}\\
&+\dfrac{\rho_{77}}{{L^5[L+2\theta(z_A+z_B-z_C)]}}
+\dfrac{\rho_{88}}{{L^5[L+2\theta(z_A+z_B+z_C)]}}\\
&+\dfrac{2}{L^6}\bigg\{ [R_{18}+R_{27}+R_{36}+R_{45}]\cos(2k\theta z_A)\cos(2k\theta
z_B)\cos(2k\theta z_C)\\
&+[I_{18}-I_{27}+I_{36}-I_{45}]\cos(2k\theta z_A)\cos(2k\theta z_B)\sin(2k\theta z_C)\\
&+[I_{18}+I_{27}-I_{36}-I_{45}]\cos(2k\theta z_A)\sin(2k\theta z_B)\cos(2k\theta z_C)\\
&+[I_{18}+I_{27}+I_{36}+I_{45}]\sin(2k\theta z_A)\cos(2k\theta z_B)\cos(2k\theta z_C)\\
&+[-R_{18}-R_{27}+R_{36}+R_{45}]\sin(2k\theta z_A)\sin(2k\theta z_B)\cos(2k\theta z_C)\\
&+[-R_{18}+R_{27}+R_{36}-R_{45}]\cos(2k\theta z_A)\sin(2k\theta z_B)\sin(2k\theta z_C)\\
&+[-I_{18}+I_{27}+I_{36}-I_{45}]\sin(2k\theta z_A)\sin(2k\theta z_B)\sin(2k\theta z_C)\\
&+[-R_{18}+R_{27}-R_{36}+R_{45}]\sin(2k\theta z_A)\cos(2k\theta z_B)\sin(2k\theta z_C)\\
&+[R_{17}+R_{28}+R_{35}+R_{46}]\cos(2k\theta z_A)\cos(2k\theta z_B) \\
&+[I_{17}+I_{28}-I_{35}-I_{46}]\cos(2k\theta z_A)\sin(2k\theta z_B) \\
&+[I_{17}+I_{28}-I_{35}+I_{46}]\sin(2k\theta z_A)\cos(2k\theta z_B) \\
&+[-R_{17}-R_{28}+R_{35}+R_{46}]\sin(2k\theta z_A)\sin(2k\theta z_B) \\
&+[R_{16}+R_{25}+R_{38}+R_{47}]\cos(2k\theta z_A)\cos(2k\theta z_C) \\
&+[I_{16}-I_{25}+I_{38}-I_{47}]\cos(2k\theta z_A)\sin(2k\theta z_C) \\
&+[I_{16}+I_{25}+I_{38}+I_{47}]\sin(2k\theta z_A)\cos(2k\theta z_C) \\
&+[-R_{16}+R_{25}-R_{38}+R_{47}]\sin(2k\theta z_A)\sin(2k\theta z_C) \\
&+[R_{14}+R_{23}+R_{58}+R_{67}]\cos(2k\theta z_B)\cos(2k\theta z_C) \\
&+[I_{14}-I_{23}+I_{58}+I_{67}]\cos(2k\theta z_B)\sin(2k\theta z_C) \\
&+[I_{14}+I_{23}+I_{58}-I_{67}]\sin(2k\theta z_B)\cos(2k\theta z_C) \\
&+[-R_{14}+R_{23}-R_{58}+R_{67}]\sin(2k\theta z_B)\sin(2k\theta z_C) \\
&+[R_{15}+R_{26}+R_{37}+R_{48}]\cos(2k\theta z_A)\\
&+[I_{15}+I_{26}+I_{37}+I_{48}]\sin(2k\theta z_A)\\
&+[R_{13}+R_{24}+R_{57}+R_{68}]\cos(2k\theta z_B)\\
&+[I_{13}+I_{24}+I_{57}+RI{68}]\sin(2k\theta z_B)\\
&+[R_{12}+R_{34}+R_{56}+R_{78}]\cos(2k\theta z_C)\\
&+[R_{12}+R_{34}+R_{56}+R_{78}]\sin(2k\theta z_C) \bigg\},
\end{aligned}
\end{eqnarray}
with $\rho_{ij}=R_{ij}+iI_{ij}$ being the coefficients of the density operator in Eq.~(\ref{rho3}), i.e., $R_{ij}$ and $I_{ij}$ are the real and imaginary parts of $\rho_{ij}$, respectively.

Notably, each of the 34 terms in Eq.~(\ref{rho31}) has an important physical significance. The first eight terms, which depend on the diagonal entries $\rho_{ii}$ of the density matrix of the operator of Eq.~(\ref{rho3}), represent the probability density of measuring particles $a$, $b$ and $c$, respectively at $z_A$, $z_B$ and $z_C$, for the cases in which there is a complete information about the slits they pass through before detection. As such, these terms alone cannot give rise to any type of interference effects. The following eight terms, which are expressed in terms of products of three oscillatory functions of $z_A$, $z_B$ or $z_C$, are the responsible for the emergence of all types of three-particle interference, i.e., interference in the CDR when all three particles are monitored. In a similar fashion, the following twelve terms, which contain products of two oscillatory functions of one of the variables $z_A$, $z_B$ and $z_C$, correspond to all possible two-particle interference patterns, that is, interference in the CDR when the particles are monitored in pairs. Finally, the last six terms, which are written in terms of single oscillatory functions of one of the variables $z_A$, $z_B$ and $z_C$, account for all types single-particle interference. These terms are the responsible for the emergence of spatial interference on the screens $S_{A}$, $S_{B}$ and $S_{C}$.  

It is well known that there exist two different classes of genuine tripartite entangled states for three-qubit systems \cite{dur,datta}. One represented by the Greenberger-Horne-Zeilinger ($GHZ$) state, $\ket{GHZ} = 1/\sqrt{2}(\ket{000}+\ket{111})$, and the other by the $W$ state, $\ket{W} = 1/\sqrt{3}(\ket{001}+\ket{010})+\ket{100})$. For the $GHZ$ state, if we measure the state of one of the subsystems such that the resulting state is either $\ket{0}$ or $\ket{1}$, the other two subsystems are left in a separable pure state. Differently, the $W$ state retains a pair of subsystems in a maximally bipartite entangled state when one of the subsystems is measured under equivalent conditions \cite{marcio}. In this context, a very important property of three-qubit systems is that states pertaining to these two different classes cannot be converted into each other by any local operation and classical communication (LOCC) process. By the same token, we infer that a three-qubit system can produce two types of three-particle interference: one observed when all three particles are monitored simultaneously, that we call $GHZ$-like interference; and another which appears only when the particles are monitored in pairs, irrespective of the pair, that we call $W$-like interference. We shall see that these two types of interference are independent of each other. In this perspective, this means that a general interference quantifier, which is supposed to work for an arbitrary three-qubit system, mixed or not, must quantify separately the amount of $GHZ$-like and $W$-like interference, and then sum both together. In this form, our first objective in this section is to understand how to quantify these two types of interference behaviors from the single-, two- and three-particle interference produced by the system of Fig. 4. To do this, we must first understand in more details what kind of interference signatures $GHZ$ and $W$ states may exhibit in that apparatus.                

Initially, let us address the issue of how $GHZ$ states manifest interference effects in the scheme of Fig. 4. In this case, the central idea is that interference in a $GHZ$ state is only detected if the three particles are jointly observed, and no quantum effect is evident when only two or one of the particles are evaluated \cite{GHZ,carvacho}. Therefore, we expect that the only interference effect produced by $GHZ$ states in the experiment of Fig. 4 is an oscillation in the CDR of the three particles, when their arrival times on the screens $S_{A}$ $S_{B}$ and $S_{C}$ are investigated. Such oscillations are observed when one varies the relative position among the detection points $z_{A}$, $z_{B}$ and $z_{C}$. On the other hand, no other type of interference effect is expected for $GHZ$ states, e.g. oscillations in the CDR when only two of the particles are observed, or single-particle spatial interference. Given this idea, our next step is to find a way to quantify genuine three-particle interference effects from the expression of the probability density in Eq.~(\ref{rho31}). As mentioned above, the terms which contain the product of three oscillatory functions of the detection points $z_{A}$, $z_{B}$ and $z_{C}$ are the responsible for the appearance of three-particle interference. However, we have to observe that combinations of the product of two oscillatory functions with a single oscillatory function, or of three single oscillatory functions, can produce a similar behavior. Therefore, to correctly extract the effect of $GHZ$-like three-particle interference, we must subtract the effect of these combinations from that of the three-particle interference to express the genuine three-particle interference. 

By inspection of Eq.~(\ref{rho31}), if we are interested for example in studying the emergence of an interference effect of the type $\cos(2k\theta y_A)\cos(2k\theta y_B)\cos(2k\theta y_C)$, we see that it can happen whether the sum $R_{18}+R_{27}+R_{36}+R_{45}$, one of the products $(R_{15}+R_{26}+R_{37}+R_{48})(R_{14}+R_{23}+R_{58}+R_{67})$,
$(R_{13}+R_{24}+R_{57}+R_{68})(R_{16}+R_{25}+R_{38}+R_{47})$,
$(R_{12}+R_{34}+R_{56}+R_{78})(R_{17}+R_{28}+R_{35}+R_{46})$, or the product $(R_{15}+R_{26}+R_{37}+R_{48})(R_{13}+R_{24}+R_{57}+R_{68})
(R_{12}+R_{34}+R_{56}+R_{78})$ is nonzero. However, similar to the case of two qubits, we have to pay attention to which of these terms really contribute with genuine three-particle interference of this type. Among all these contributions, the only one which embodies the possibility of genuine three-particle interference is the sum $R_{18}+R_{27}+R_{36}+R_{45}$. Besides the interference mode of the type $\cos(2k\theta y_A)\cos(2k\theta y_B)\cos(2k\theta y_C)$, seven other types of three-particle interference, which are LI, can take place. They will all be listed below. Nevertheless, before doing that, we anticipate that our $GHZ$-like interference quantifier will be given by the imbalance among the three-, two- and single-particle interference contributions to the CDR of all eight LI oscillatory modes. Given this phenomenological definition, the amount of $GHZ$-like interference contained in each of the eight possible three-particle interference modes are given by:  
\\
\newline
i) $\cos(2k\theta y_A)\cos(2k\theta y_B)\cos(2k\theta y_C)$ mode for particles A-B-C:
\begin{eqnarray}
\begin{aligned}
\label{GHZ1}
\mathcal{I}_{GHZ}^{(1)}=&4|(R_{18}+R_{27}+R_{36}+R_{45})^2- \\
&4[(R_{15}+R_{26}+R_{37}+R_{48})(R_{14}+R_{23}+R_{58}+R_{67})+\\
&(R_{13}+R_{24}+R_{57}+R_{68})(R_{16}+R_{25}+R_{38}+R_{47})+\\
&(R_{12}+R_{34}+R_{56}+R_{78})(R_{17}+R_{28}+R_{35}+R_{46})]^2-\\
&16[(R_{15}+R_{26}+R_{37}+R_{48})(R_{13}+R_{24}+R_{57}+R_{68})
(R_{12}+R_{34}+R_{56}+R_{78})]^2|.\\
\end{aligned}
\end{eqnarray}
ii) $\cos(2k\theta y_A)\cos(2k\theta y_B)\sin(2k\theta y_C)$ mode  for particles A-B-C:
\begin{eqnarray}
\begin{aligned}
\label{GHZ2}
\mathcal{I}_{GHZ}^{(2)}=&4|(I_{18}-I_{27}+I_{36}-I_{45})^2- \\
&4[(R_{15}+R_{26}+R_{37}+R_{48})(I_{14}+I_{23}+I_{58}+I_{67})+\\
&(R_{13}+R_{24}+R_{57}+R_{68})(I_{16}-I_{25}+I_{38}-I_{47})+\\
&(I_{12}+I_{34}+I_{56}+I_{78})(R_{17}+R_{28}+R_{35}+R_{46})]^2-\\
&16[(R_{15}+R_{26}+R_{37}+R_{48})(R_{13}+R_{24}+R_{57}+R_{68})
(I_{12}+I_{34}+I_{56}+I_{78})]^2|.\\
\end{aligned}
\end{eqnarray}
iii) $\cos(2k\theta y_A)\sin(2k\theta y_B)\cos(2k\theta y_C)$ mode  for particles A-B-C:
\begin{eqnarray}
\begin{aligned}
\label{GHZ3}
\mathcal{I}_{GHZ}^{(3)}=&4|(I_{18}+I_{27}-I_{36}-I_{45})^2-\\
&4[(R_{15}+R_{26}+R_{37}+R_{48})(I_{14}-I_{23}+I_{58}-I_{67})+\\
&(I_{13}+I_{24}+I_{57}+I_{68})(R_{16}+R_{25}+R_{38}+R_{47})+\\
&(R_{12}+R_{34}+R_{56}+R_{78})(I_{17}+I_{28}-I_{35}-I_{46})]^2-\\
&16[(R_{15}+R_{26}+R_{37}+R_{48})(I_{13}+I_{24}+I_{57}+I_{68})
(R_{12}+R_{34}+R_{56}+R_{78})]^2|.\\
\end{aligned}
\end{eqnarray}
iv) $\sin(2k\theta y_A)\cos(2k\theta y_B)\cos(2k\theta y_C)$ mode  for particles A-B-C:
\begin{eqnarray}
\begin{aligned}
\label{GHZ4}
\mathcal{I}_{GHZ}^{(4)}=&4|(I_{18}+I_{27}+I_{36}+I_{45})^2-\\
&4[(I_{15}+I_{26}+I_{37}+I_{48})(R_{14}+R_{23}+R_{58}+R_{67})+\\
&(R_{13}+R_{24}+R_{57}+R_{68})(I_{16}+I_{25}+I_{38}+I_{47})+\\
&(R_{12}+R_{34}+R_{56}+R_{78})(I_{17}+I_{28}-I_{35}+I_{46})]^2-\\
&16[(I_{15}+I_{26}+I_{37}+I_{48})(R_{13}+R_{24}+R_{57}+R_{68})
(R_{12}+R_{34}+R_{56}+R_{78})]^2|.\\
\end{aligned}
\end{eqnarray}
v) $\sin(2k\theta y_A)\sin(2k\theta y_B)\cos(2k\theta y_C)$ mode  for particles A-B-C:
\begin{eqnarray}
\begin{aligned}
\label{GHZ5}
\mathcal{I}_{GHZ}^{(5)}=&4|(-R_{18}-R_{27}+R_{36}+R_{45})^2-\\
&4[(I_{15}+I_{26}+I_{37}+I_{48})(I_{14}-I_{23}+I_{58}-I_{67})+\\
&(I_{13}+I_{24}+I_{57}+I_{68})(I_{16}+I_{25}+I_{38}+I_{47})+\\
&(R_{12}+R_{34}+R_{56}+R_{78})(-R_{17}-R_{28}+R_{35}+R_{46})]^2-\\
&16[(I_{15}+I_{26}+I_{37}+I_{48})(I_{13}+I_{24}+I_{57}+I_{68})
(R_{12}+R_{34}+R_{56}+R_{78})]^2|.\\
\end{aligned}
\end{eqnarray}
vi) $\cos(2k\theta y_A)\sin(2k\theta y_B)\sin(2k\theta y_C)$ mode  for particles A-B-C:
\begin{eqnarray}
\begin{aligned}
\label{GHZ6}
\mathcal{I}_{GHZ}^{(6)}=&4|(-R_{18}+R_{27}+R_{36}-R_{45})^2-\\
&4[(R_{15}+R_{26}+R_{37}+R_{48})(-R_{14}+R_{23}-R_{58}+R_{67})+\\
&(I_{13}+I_{24}+I_{57}+I_{68})(I_{16}-I_{25}+I_{38}-I_{47})+\\
&(I_{12}+I_{34}+I_{56}+I_{78})(I_{17}+I_{28}-I_{35}-I_{46})]^2-\\
&16[(R_{15}+R_{26}+R_{37}+R_{48})(I_{13}+I_{24}+I_{57}+I_{68})
(I_{12}+I_{34}+I_{56}+I_{78})]^2|.\\
\end{aligned}
\end{eqnarray}
vii) $\sin(2k\theta y_A)\sin(2k\theta y_B)\sin(2k\theta y_C)$ mode  for particles A-B-C:
\begin{eqnarray}
\begin{aligned}
\label{GHZ7}
\mathcal{I}_{GHZ}^{(7)}=&4|(-I_{18}+I_{27}+I_{36}-I_{45})^2- \\
&4[(I_{15}+I_{26}+I_{37}+I_{48})(-R_{14}+R_{23}-R_{58}+R_{67})+\\
&(I_{13}+I_{24}+I_{57}+I_{68})(-R_{16}+R_{25}+R_{38}+R_{47})+\\
&(I_{12}+I_{34}+I_{56}+I_{78})(-R_{17}-R_{28}-R_{35}-R_{46})]^2-\\
&16[(I_{15}+I_{26}+I_{37}+I_{48})(I_{13}+I_{24}+I_{57}+I_{68})
(I_{12}+I_{34}+I_{56}+I_{78})]^2|.\\
\end{aligned}
\end{eqnarray}
viii) $\sin(2k\theta y_A)\cos(2k\theta y_B)\sin(2k\theta y_C)$ mode  for particles A-B-C:
\begin{eqnarray}
\begin{aligned}
\label{GHZ8}
\mathcal{I}_{GHZ}^{(8)}=&4|(-R_{18}+R_{27}-R_{36}+R_{45})^2-\\
&4[(I_{15}+I_{26}+I_{37}+I_{48})(I_{14}-I_{23}+I_{58}+I_{67})+\\
&(R_{13}+R_{24}+R_{57}+R_{68})(-R_{16}+R_{25}-R_{38}+R_{47})+\\
&(I_{12}+I_{34}+I_{56}+I_{78})(I_{17}+I_{28}-I_{35}+I_{46})]^2-\\
&16[(I_{15}+I_{26}+I_{37}+I_{48})(R_{13}+R_{24}+R_{57}+R_{68})
(I_{12}+I_{34}+I_{56}+I_{78})]^2|.\\
\end{aligned}
\end{eqnarray}

As can be seen, the eight interference mode quantifiers represent the imbalance among the square of the three-, two- and single interference contributions to each of the possible LI modes, similar to what was realized in the two-qubit case. If such imbalance is zero, it means that the oscillatory mode in question is not detectable. Therefore, if the tripartite state has some amount of genuine $GHZ$-like interference, at least one of the above mode quantifiers will be nonzero. Overall, the total amount of $GHZ$-like interference in an arbitrary three-qubit state is given by 
\begin{eqnarray} 
\label{EGHZ}
\mathcal{I}_{GHZ}(\hat{\rho}) = \frac{1}{4}\sum_{i=1}^{8} \mathcal{I}_{GHZ}^{(i)}, 
\end{eqnarray}
with the $\mathcal{I}_{GHZ}^{(i)}$ elements given in Eqs.~(\ref{GHZ1}) to~(\ref{GHZ8}). The multiplicative (normalization) factor $1/4$ was placed in order to obtain $\mathcal{I}_{GHZ}(\hat{\rho}) =1$ for the maximally entangled $GHZ$ state.

Let us now give some examples to illustrate the validity of the interference quantifier of Eq.~(\ref{EGHZ}). Again, for clarity of the examples, we will use the computational basis in the following form: $\ket{A_{1}}\ket{B_{1}}\ket{C_{1}} \rightarrow \ket{000}$; $\ket{A_{1}}\ket{B_{1}}\ket{C_{2}} \rightarrow \ket{001}$; $\ket{A_{1}}\ket{B_{2}}\ket{C_{1}} \rightarrow \ket{010}$; $\ket{A_{1}}\ket{B_{2}}\ket{C_{2}} \rightarrow \ket{011}$; $\ket{A_{2}}\ket{B_{1}}\ket{C_{1}} \rightarrow \ket{100}$; $\ket{A_{2}}\ket{B_{1}}\ket{C_{2}} \rightarrow \ket{101}$; $\ket{A_{2}}\ket{B_{2}}\ket{C_{1}} \rightarrow \ket{110}$ and $\ket{A_{2}}\ket{B_{2}}\ket{C_{2}} \rightarrow \ket{111}$. In this form, as a first example we have that the state $\ket{\psi}_1=\cos(\alpha)\ket{000}+e^{i\phi}\sin(\alpha)\ket{111}$ provides $\mathcal{I}_{GHZ}=\sin^2(2\alpha)$, which attains unit only for the maximally entangled $GHZ$ state, when $\alpha = \pi/4$ or $3 \pi/4$. On the other hand, for the important case of a general pure separable state, $\ket{\psi}_{2} = \ket{\psi_{A}}\ket{\psi_{B}}\ket{\psi_{C}}$, with $\ket {\psi_{A}}=\cos (\theta_1 /2)\ket{0}+e^{i\phi_1}\sin (\theta_1 /2)\ket{1}$, $\ket{\psi_{B}}=\cos (\theta_2 /2)\ket{0}+e^{i\phi_2}\sin (\theta_2 /2) \ket{1}$ and $\ket{\psi_{C}}=\cos (\theta_3 /2)\ket{0}+e^{i\phi_3}\sin (\theta_3 /2) \ket{1}$, after some calculations we find that $\mathcal{I}_{GHZ}=0$.

Now, we address the issue of how $W$ states produce interference effects in the apparatus of Fig. 4. In this case, we have that the three particles must manifest interference in the CDR when they are observed two at a time, but no interference is detected when the three particles are simultaneously monitored, and no single-particle (spatial) interference is visualized. Thus, interference effects for these states are only detectable for the arrival times of the particles when they are measured in pairs. That is to say that oscillations can be found only when one varies the relative position between any pair of detection points among $z_{A}$, $z_{B}$ and $z_{C}$. As such, we now proceed to find a way to quantify $W$-like interference from the probability density given in Eq.~(\ref{rho31}). In doing so, we first need to identify all terms which contain the product of two oscillatory functions of the variables $z_{A}$, $z_{B}$ and $z_{C}$, as well as the combination (product) of two single oscillatory functions producing similar effects. Below, we list the twelve possible LI oscillatory modes which are important for quantifying $W$-like interference, and the corresponding contributions that they have from the terms of Eq.~(\ref{rho31}). Again, we shall consider the contributions of two-particle interference and subtract the analogous contribution due to combinations of single oscillatory functions in order to express the genuine two-particle interferences. As we have done in the case of two-qubits and $GHZ$ states, we quantify the amount of interference for each mode with the absolute value of the difference between the square of the two-particle and combined contributions:       
\\               
\newline
i) $\cos(2k\theta y_A)\cos(2k\theta y_B)$ mode  for particles A-B:
\begin{equation}
\label{Wab1}
\mathcal{I}_{Wab}^{(1)} = 4|(R_{17}+R_{28}+R_{35}+R_{46})^2-4[(R_{15}+R_{26}+R_{37}+R_{48})
(R_{13}+R_{24}+R_{57}+R_{68})]^2|.
\end{equation}
\\
ii) $\cos(2k\theta y_A)\sin(2k\theta y_B)$ mode for particles A-B:
\begin{equation}
\mathcal{I}_{Wab}^{(2)} =4|(I_{17}+I_{28}-I_{35}-I_{46})^2-4[(R_{15}+R_{26}+R_{37}+R_{48})
(I_{13}+I_{24}+I_{57}+I_{68})]^2|.
\end{equation} \newline
iii) $\sin(2k\theta y_A)\cos(2k\theta y_B)$ mode for particles A-B:
\begin{equation}
\mathcal{I}_{Wab}^{(3)}= 4|(I_{17}+I_{28}-I_{35}+I_{46})^2-4[(I_{15}+I_{26}+I_{37}+I_{48})
(R_{13}+R_{24}+R_{57}+R_{68})]^2|.
\end{equation} \newline
iv) $\sin(2k\theta y_A)\sin(2k\theta y_B)$ mode for particles A-B:
\begin{equation}
\mathcal{I}_{Wab}^{(4)}= 4|(-R_{17}-R_{28}+R_{35}+R_{46})^2-4[(I_{15}+I_{26}+I_{37}+I_{48})
(I_{13}+I_{24}+I_{57}+I_{68})]^2|.
\end{equation} \newline
v) $\cos(2k\theta y_A)\cos(2k\theta y_C)$ mode for particles A-C:
\begin{equation}
\mathcal{I}_{Wac}^{(1)} =4|(R_{16}+R_{25}+R_{38}+R_{47})^2-4[(R_{15}+R_{26}+R_{37}+R_{48})
(R_{12}+R_{34}+R_{56}+R_{78})]^2|.
\end{equation} \newline
vi) $\cos(2k\theta y_A)\sin(2k\theta y_C)$ mode for particles A-C:
\begin{equation}
\mathcal{I}_{Wac}^{(2)}= 4|(I_{16}-I_{25}+I_{38}-I_{47})^2-4[(R_{15}+R_{26}+R_{37}+R_{48})
(I_{12}+I_{34}+I_{56}+I_{78})]^2|.
\end{equation} \newline
vii) $\sin(2k\theta y_A)\cos(2k\theta y_C)$ mode for particles A-C:
\begin{equation}
\mathcal{I}_{Wac}^{(3)}= 4|(I_{16}+I_{25}+I_{38}+I_{47})^2-4[(I_{15}+I_{26}+I_{37}+I_{48})
(R_{12}+R_{34}+R_{56}+R_{78})]^2|.
\end{equation} \newline
viii) $\sin(2k\theta y_A)\sin(2k\theta y_C)$ mode for particles A-C:
\begin{equation}
\mathcal{I}_{Wac}^{(4)}= 4|(-R_{16}+R_{25}-R_{38}+R_{47})^2-4[(I_{15}+I_{26}+I_{37}+I_{48})
(I_{12}+I_{34}+I_{56}+I_{78})]^2|.
\end{equation} \newline
ix) $\cos(2k\theta y_B)\cos(2k\theta y_C)$ mode for particles B-C:
\begin{equation}
\mathcal{I}_{Wbc}^{(1)}= 4|(R_{14}+R_{23}+R_{58}+R_{67})^2-4[(R_{13}+R_{24}+R_{57}+R_{68})
(R_{12}+R_{34}+R_{56}+R_{78})]^2|.
\end{equation} \newline
x) $\cos(2k\theta y_B)\sin(2k\theta y_C)$ mode for particles B-C:
\begin{equation}
\mathcal{I}_{Wbc}^{(2)}= 4|(I_{14}-I_{23}+I_{58}+I_{67})^2-4[(R_{13}+R_{24}+R_{57}+R_{68})
(I_{12}+I_{34}+I_{56}+I_{78})]^2|.
\end{equation} \newline
xi) $\sin(2k\theta y_B)\cos(2k\theta y_C)$ mode for particles B-C:
\begin{equation}
\mathcal{I}_{Wbc}^{(3)}=4|(I_{14}+I_{23}+I_{58}-I_{67})^2-4[(I_{13}+I_{24}+I_{57}+I_{68})
(R_{12}+R_{34}+R_{56}+R_{78})]^2|.
\end{equation} \newline
xii) $\sin(2k\theta y_B)\sin(2k\theta y_C)$ mode for particles B-C:
\begin{equation}
\label{Wbc4}
\mathcal{I}_{Wbc}^{(4)}= 4|(-R_{14}+R_{23}-R_{58}+R_{67})^2-4[(I_{13}+I_{24}+I_{57}+I_{68})
(I_{12}+I_{34}+I_{56}+I_{78})]^2|.
\end{equation} \newline

Given these quantifiers for all possible two-particle interference modes, and being aware that a $W$-like interference only exists if all three particles are quantum correlated, but with the quantum correlations established only for pairs of particles, we will now write the interference quantifier for $W$-like states. To this end, we must understand that for the $W$-like interference to exist, at least one of the modes of each type A-B, A-C, and B-C must be nonzero. Therefore, the total amount of $W$-like interference in an arbitrary three-qubit state is given by    
 
\begin{eqnarray} 
\label{EW}
\mathcal{I}_{W}(\hat{\rho}) = \left(\frac{9}{8} \right)^{3} \left[\sum_{i=1}^{4} \mathcal{I}_{Wab}^{(i)}\sum_{j=1}^{4}\mathcal{I}_{Wac}^{(j)} \sum_{k=1}^{4} \mathcal{I}_{Wbc}^{(k)}\right], 
\end{eqnarray}
with the $\mathcal{I}_{Wab}^{(i)}$, $\mathcal{I}_{Wac}^{(i)}$ and $\mathcal{I}_{Wbc}^{(k)}$ elements given as in Eqs.~(\ref{Wab1}) to~(\ref{Wbc4}). The normalization factor $(9/8)^{3}$ was introduced to set $\mathcal{I}_{W}(\hat{\rho}) = 1$ for the maximally entangled $W$ state. By testing this quantifier for an arbitrary pure separable state $\ket{\psi}_{1} = \ket{\psi_{A}}\ket{\psi_{B}}\ket{\psi_{C}}$, with $\ket {\psi_{A}}=\cos (\theta_1 /2)\ket{0}+e^{i\phi_1}\sin (\theta_1 /2)\ket{1}$, $\ket{\psi_{B}}=\cos (\theta_2 /2)\ket{0}+e^{i\phi_2}\sin (\theta_2 /2) \ket{1}$ and $\ket{\psi_{C}}=\cos (\theta_3 /2)\ket{0}+e^{i\phi_3}\sin (\theta_3 /2) \ket{1}$, we obtain $\mathcal{I}_{W}(\hat{\rho}) = 0$, as expected. Also, for a general $W$ state, $\ket{\psi}_2=\dfrac{1}{\sqrt{3}}[\ket{100}+e^{i\phi_1}\ket{010}+e^{i\phi_2}\ket{001}]$, we obtain $\mathcal{I}_{W}(\hat{\rho}) = 1$. 

Overall, if we want to quantify interference for a general three-qubit state, we just need to sum the quantifiers for $GHZ$-like and $W$-like states, which are independent of each other, i.e., quantify different three-qubit classes of multiparticle interference. In this form, the general quantifier is:
\begin{eqnarray} 
\label{final}
\mathcal{I}^{(2)}_{3}(\hat{\rho}) = \mathcal{I}_{GHZ}(\hat{\rho}) + \mathcal{I}_{W}(\hat{\rho}),
\end{eqnarray}
with $\mathcal{I}_{GHZ}(\hat{\rho})$ and $\mathcal{I}_{W}(\hat{\rho})$ given by Eqs.~(\ref{EGHZ}) and~(\ref{EW}), respectively. The subscript 3 and superscript 2 in $\mathcal{I}^{(2)}_{3}(\hat{\rho})$ stand for the number of parties and the dimension of the Hilbert space of each party, respectively. According to the tests realized with this general interference quantifier, we observed that $\mathcal{I}_{GHZ}(\hat{\rho})>0$ and $\mathcal{I}_{W}(\hat{\rho}) = 0$ for all $GHZ$-like states, and $\mathcal{I}_{GHZ}(\hat{\rho})=0$ and $\mathcal{I}_{W}(\hat{\rho}) >0$ for all $W$-like states, as expected. In this respect, we can say that states that satisfy $\mathcal{I}_{GHZ}(\hat{\rho})>0$ and $\mathcal{I}_{W}(\hat{\rho}) >0$ manifest genuine $GHZ$- and $W$-like quantum interference, respectively. It is noteworthy to point out that Eqs.~(\ref{EGHZ}) and~(\ref{EW}) are quantifiers that measure respectively the amount of $GHZ$- and $W$-like quantum interference for any three-qubit state. Also, since these are the only two types of genuine tripartite quantum interferences, when summed together as in Eq.~(\ref{final}), the relation represents a state-independent quantifier for three-qubit quantum interference.

\begin{figure}[ht]
\centerline{\includegraphics[width=10.2cm]{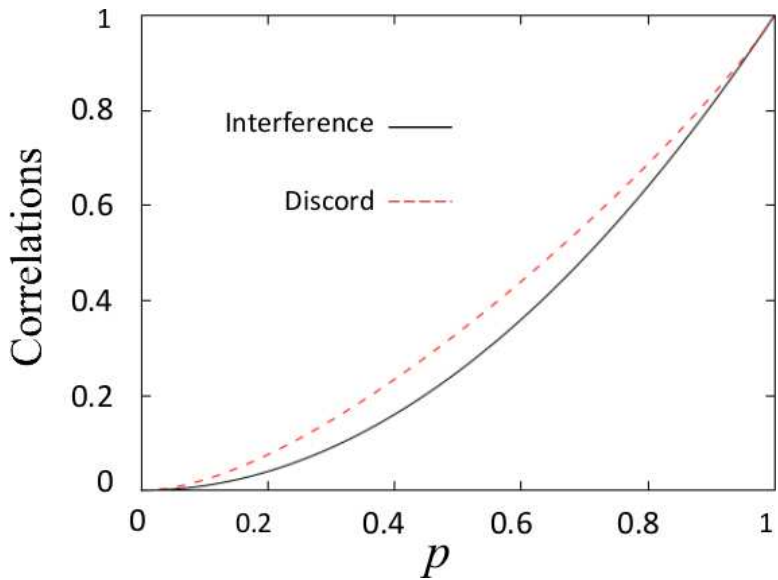}}
\caption{(Color online) Graphs of the quantum interference $\mathcal{I}_3^{(2)}(\hat{\rho}_{W}^{(GHZ)})$ and global discord for the  Werner-$GHZ$
state of Eq.~(\ref{wernerghz}). As can be seen, both quantifiers provide nonzero values in the interval $0 \leq p \leq 1/5$, in which the state is separable \cite{dur,pitt}.}
\end{figure}

As an example of our quantum interference quantifier applied to three-qubit mixed states, we consider the case of the Werner-$GHZ$
state.  This state is defined as a mixture of a $GHZ$ state and completely
depolarized noise \cite{marcio,sie,elt}:
\begin{equation}
\label{wernerghz}
 \hat{\rho}_{W}^{(GHZ)} = p \ket{GHZ}\bra{GHZ} + \frac{1-p}{8} \hat{{\bf 1}},
\end{equation}
with $0 \leq p \leq 1 $. In Fig. 5 we compare our results of quantum interference using Eq.~(\ref{final}) with those of global quantum discord computed in Ref. \cite{rulli} for this state. As can be seen, both quantum interference and discord are nonzero for $p>0$. However, this state is known to be separable if $p<1/5$ \cite{dur,pitt}, and biseparable if $p<3/7$ \cite{guh}. As such, this finding confirms that some separable states, as well as biseparable states, can exhibit three-particle interference. It is important to call attention to the fact that the quantum interference of the Werner-$GHZ$ state,   $\mathcal{I}_3^{(2)}(\hat{\rho}_{W}^{(GHZ)})$, receives exclusive $GHZ$-like contributions, i.e., $\mathcal{I}_{W}(\hat{\rho}_{W}^{(GHZ)}) = 0$ for all values of the parameter $p$.

\section{Conclusions and remarks}

In conclusion, we have presented a new method of quantifying quantum correlations based on the multiparticle interference produced by a composite quantum system when each of the constituent subsystems is submitted to a double-slit apparatus. By means of the expressions of the single- and multiparticle interference patterns manifested in this scheme, we were able to write two formulas that quantifies the amount of quantum correlation for general two- and three-qubit systems. Remarkably, these expressions could be derived explicitly in terms of the density matrix elements, independent if the bipartite or tripartite state is pure or mixed. Interestingly, for the special case of pure states, we verified that our quantum interference quantifiers for two- and three-qubit systems could also work as entanglement quantifiers. In fact, applying Eqs.~(\ref{quant}) and~(\ref{final}) to pure two- and three-qubit  states, respectively, we verified that our quantifiers, $\mathcal{I}^{(2)}_{2}(\hat{\rho})$ and $\mathcal{I}^{(2)}_{3}(\hat{\rho})$, satisfy the basic entanglement quantifier postulates \cite{horo,friis}: (i) Monotonicity under local operations and classical communication (LOCC): The amount of interference cannot increase under LOCC. (ii) $\mathcal{I}^{(2)}_{2}(\hat{\rho})  = 0$ and $ \mathcal{I}^{(2)}_{3}(\hat{\rho}) = 0$ if $\hat{\rho}$ is a separable state: we verified this point by analyzing separable pure states in both cases. Moreover, despite not being a requirement, our quantifiers also satisfy the criterion of (iii) Normalization: $\mathcal{I}(\ket{\Phi_{d}}) = \log_{2}  d$, where $\ket{\Phi_{d}}$ is a maximally entangled state, with $d$ as the dimension of the Hilbert space of each subsystem. 

On the other hand, by investigating our two quantifiers for the case of mixed Werner states, we observed that the amount of multiparticle quantum interference of some separable states is nonzero. It is important to call attention that these states were also demonstrated to have nonzero discord \cite{ollivier, hend,rulli}. In this context, from the analyses realized with two- and three-qubit Werner states, one can see that the profiles of quantum discord and our definition of quantum interference are similar, principally in the first case, as can be seen in Fig 3. Notably, our results showed that entanglement is not a requirement for a composite quantum system to produce multiparticle interference, as widely accepted in the literature \cite{green,hsz1,hsz2,jwp}. We believe the present method of quantifying quantum correlations through the study of the interference patterns produced by many-particle systems will advance our understanding of quantumness in composite systems and bring important insights
into some central questions on the foundations of quantum theory. Particularly, we envisage that our approach to quantify quantum correlations can be extended to more than three qubits, or to higher dimensional systems.

\section*{Acknowledgement}
        
One of the authors (R.A.B) has been supported by CNPq. B.L.B acknowledges financial support from the Brazilian funding agencies CAPES/Finance Code 001; CNPq, Grant Number 303451/2019-0; and Pronex/Fapesq-PB/CNPq, Grant Number 0016/2019.


\end{document}